\documentclass{PoS}

\title{Analysis of the scalar mesons on the Lattice }

\ShortTitle{Analysis of the scalar mesons on the Lattice }

\author{\speaker{Masayuki Wakayama}
\\
        Department of Physics, Nagoya University, Furo-cho, Chikusa-ku, Nagoya 464-8602, Japan\\
        E-mail: \email{wakayama@hken.phys.nagoya-u.ac.jp}}
\author{Chiho Nonaka\\
        Department of Physics, Kobayashi-Maskawa Institute for the Origin of Particles and the Universe (KMI), 
        Nagoya University, Furo-cho, Chikusa-ku, Nagoya 464-8602, Japan\\
        E-mail: \email{nonaka@hken.phys.nagoya-u.ac.jp}}

\abstract{
We study the possibility that the scalar mesons exist as four-quark states.
The energy shift of two pseudoscalar mesons as a function of spatial lattice size makes a distinction 
between bound states and scattering states of four-quark states.
We calculate the four-quark state in the quenched approximation, 
ignoring the two-quark annihilation diagrams and the vacuum channels.
We perform a calculation of pseudoscalar meson scattering amplitudes, 
using $N_f=2$ Wilson fermion and plaquette/Iwasaki gauge actions.
We obtain the indication that 
the four-quark states in the case of the isospin zero ($I=0$) and two ($I=2$) channels are no bound states.
And we find that the bound energy depends strongly on pion mass 
rather than the ratio of pion mass to rho meson mass.
}

\FullConference{The 30th International Symposium on Lattice Field Theory\\
                 June 24 - 29,  2012\\
                 Cairns, Australia}

\begin{document}

\section{Introduction}
Since many light scalar mesons such as $\sigma (600), \kappa (800), f_0(980)$ and $a_0(980)$ 
were found in experiments \cite{PDG}, a lot of theoretical studies have been devoted to 
investigation of their states.  
For the structure of light salar mesons, in addition to the conventional two quark state from the quark model, 
several possibilities are proposed \cite{Jaffe1}; four-quark states, molecular states and scattering states.
Because the sigma meson is considered as a chiral partner of the pion in the mechanism of hadron mass generation,  
it would be interesting to investigate a roll of four-quark states in the mechanism. 
The study of four-quark states in light scalar mesons gives us insight of important QCD feature.

Since Alford and Jaffe showed the possibility 
that scalar mesons exist as four-quark states on the lattice \cite{Jaffe}, 
tetraquark search on the lattice started actively.
The pioneer work of the sigma meson which is one of candidates of four-quark states 
was done with full QCD by SCALAR collaboration \cite{Scalar}.
They found that disconnected diagrams which contain effectively four-quark states, 
glueballs and so on make the sigma meson lighter. 
Recently using tetraquark interpolators, 
not only ground states but also resonance states of scalar mesons on the lattice were reported \cite{BGR}.

Following the procedure which was proposed by Alford and Jaffe \cite{Jaffe}, 
we explore the existence of four-quark states in the scalar channel on larger lattice with a finer lattice spacing.
We also investigate the dependence of bound energies for four-quark states 
on the ratio of pion mass and rho meson mass ($m_{\pi}/m_{\rho}$)  and on the pion mass ($m_{\pi}$).

\begin{figure}[b]
\vspace{0.35cm}
\centering
\begin{picture}(250,50)
 \qbezier(-90,32)(-40,82)(10,32)
 \qbezier(-90,32)(-40,52)(10,32)
 \qbezier(-90,28)(-40,8)(10,28)
 \qbezier(-90,28)(-40,-22)(10,28)
\put(-38,57){\vector(1,0){0.1}}
\put(-42,42){\vector(-1,0){0.1}}
\put(-38,18){\vector(1,0){0.1}}
\put(-42,3){\vector(-1,0){0.1}}

 \qbezier(20,32)(70,82)(120,32)
 \qbezier(20,32)(33,38)(51,41)
 \qbezier(89,41)(107,38)(120,32)
 \qbezier(20,28)(33,22)(51,19)
 \qbezier(89,19)(107,22)(120,28)
 \qbezier(20,28)(70,-22)(120,28)
\put(51.6,41){\line(5,-3){37}}
\put(51.6,19){\line(5,3){17}}
\put(88.6,41){\line(-5,-3){17}}
\put(72,57){\vector(1,0){0.1}}
\put(76,33.2){\vector(-4,-3){0.1}}
\put(58,22.8){\vector(-4,-3){0.1}}
\put(76,26.4){\vector(-4,3){0.1}}
\put(58,37.2){\vector(-4,3){0.1}}
\put(72,3){\vector(1,0){0.1}}

 \qbezier(130,32)(180,82)(230,32)
 \qbezier(130,28)(180,-22)(230,28)
\put(182,57){\vector(1,0){0.1}}
\put(178,3){\vector(-1,0){0.1}}
 \qbezier(130,32)(158,50)(160,30)
 \qbezier(130,28)(158,10)(160,30)
 \qbezier(200,30)(202,50)(230,32)
 \qbezier(200,30)(202,10)(230,28)
\put(160,32){\vector(0,1){0.1}}
\put(200,28){\vector(0,-1){0.1}}

 \qbezier(240,32)(281,72)(285,30)
 \qbezier(240,28)(281,-12)(285,30)
 \qbezier(295,30)(299,72)(340,32)
 \qbezier(295,30)(299,-12)(340,28)
\put(285,28){\vector(0,-1){0.1}}
\put(295,32){\vector(0,1){0.1}}
 \qbezier(240,32)(268,50)(270,30)
 \qbezier(240,28)(268,10)(270,30)
 \qbezier(310,30)(312,50)(340,32)
 \qbezier(310,30)(312,10)(340,28)
\put(270,32){\vector(0,1){0.1}}
\put(310,28){\vector(0,-1){0.1}}

 \put(-91,52){$D(t)$}
 \put(20,52){$C(t)$}
 \put(130,52){$A(t)$}
 \put(238,52){$G(t)$}

\end{picture}
\caption{The diagrams for four-quark correlators.}
\label{DCAG}
\end{figure}
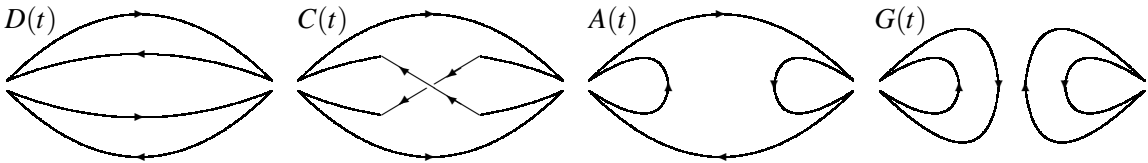

\section{Four-quark states from Lattice QCD}
We calculate four-quark correlators in two-flavor ($N_f=2$) lattice QCD 
under the assumption that four-quark states of scalar mesons are consisted of bound states of the two pions.
We employ the procedure which was proposed by Alford and Jaffe \cite{Jaffe}. 
The four-quark correlators of isospin zero ($I=0$) and two ($I=2$) channels are given by, 
\begin{eqnarray}
J_{I=0}(t) &=& D(t)+\frac{1}{2}C(t) -3A(t) + \frac{3}{2}G(t) \ , \\
J_{I=2}(t) &=& D(t)-C(t) \ ,
\end{eqnarray} 
where $D(t)$, $C(t)$, $A(t)$ and $G(t)$ are corresponding to the diagrams in Fig.\ \ref{DCAG}.
To evaluate the four-quark states clearly, 
we carry out the calculation in the quenched  approximation 
where two-quark, multi-quark and glueball states do not mix with four-quark states as intermediate states.
We drop the contribution of the diagrams for $A(t)$ and $G(t)$, 
ignoring the two-quark annihilation diagrams and the vacuum channels.
To obtain the bound energy $\delta E_{I}$ for the four-quark states, 
we construct the ratio of the four-quark correlator $J_I(t)$ and the pion correlator $P(t)$ 
and fit it to an exponential at large $t$, 
\begin{eqnarray}
 R_{I}(t) \ =\                                      \frac{J_{I}(t)}{\left(P(t)\right)^2} 
             &\longrightarrow&               \frac{Z_{I}}{Z_{\pi}^2} \exp{(-\delta E_{I}\: t)} + \cdots \ . \\[-0.45cm]
             &{\scriptstyle {t \to \infty}}& \nonumber 
\end{eqnarray}
We discuss the possibility that the four-quark states exist as bound states 
from the $N_L$ dependence of the bound energies.

If a four-quark state is a bound state, 
$\delta E_{I}$ is negative and 
would approach to a negative constant in the large $N_L$ region.
On the other hand 
if a four-quark state is a scattering state, 
it is expected that $\delta E_{I}$ obeys the scattering formula of a two-particle state 
in a cube box of size $L$ with periodic boundary condition in Ref.\ \cite{Luscher}, 
\begin{eqnarray}
 \delta E_{I} 
 = E_{I} - 2m_{\pi}
 = \frac{T_{I}}{L^3} \left[ 1 + 2.8373 \left(\frac{m_{\pi}T_{I}}{4\pi L}  \right)
                                         + 6.3752 \left(\frac{m_{\pi}T_{I}}{4\pi L} \right)^2  \right] + {\cal{O}} \left(L^{-6}\right) \ ,
\label{Luscher}
\end{eqnarray}
where $E_{I}$ is the total energy, and $T_{I}$ is the scattering amplitude 
which can be written by the scattering length $a_I$ as 
\begin{eqnarray}
  T_{I} = - \frac{4\pi a_I}{m_{\pi}} \ .
\label{amp}
\end{eqnarray}
From Eq.(\ref{Luscher}), we find that  
$\delta E_{I}$ is proportional to $L^{-3}$ in the region 
where the physical spatial lattice size $L=N_L a$ ($a$ is the lattice spacing) is enough large.

\begin{figure}
\begin{center}
\includegraphics[scale=0.455]{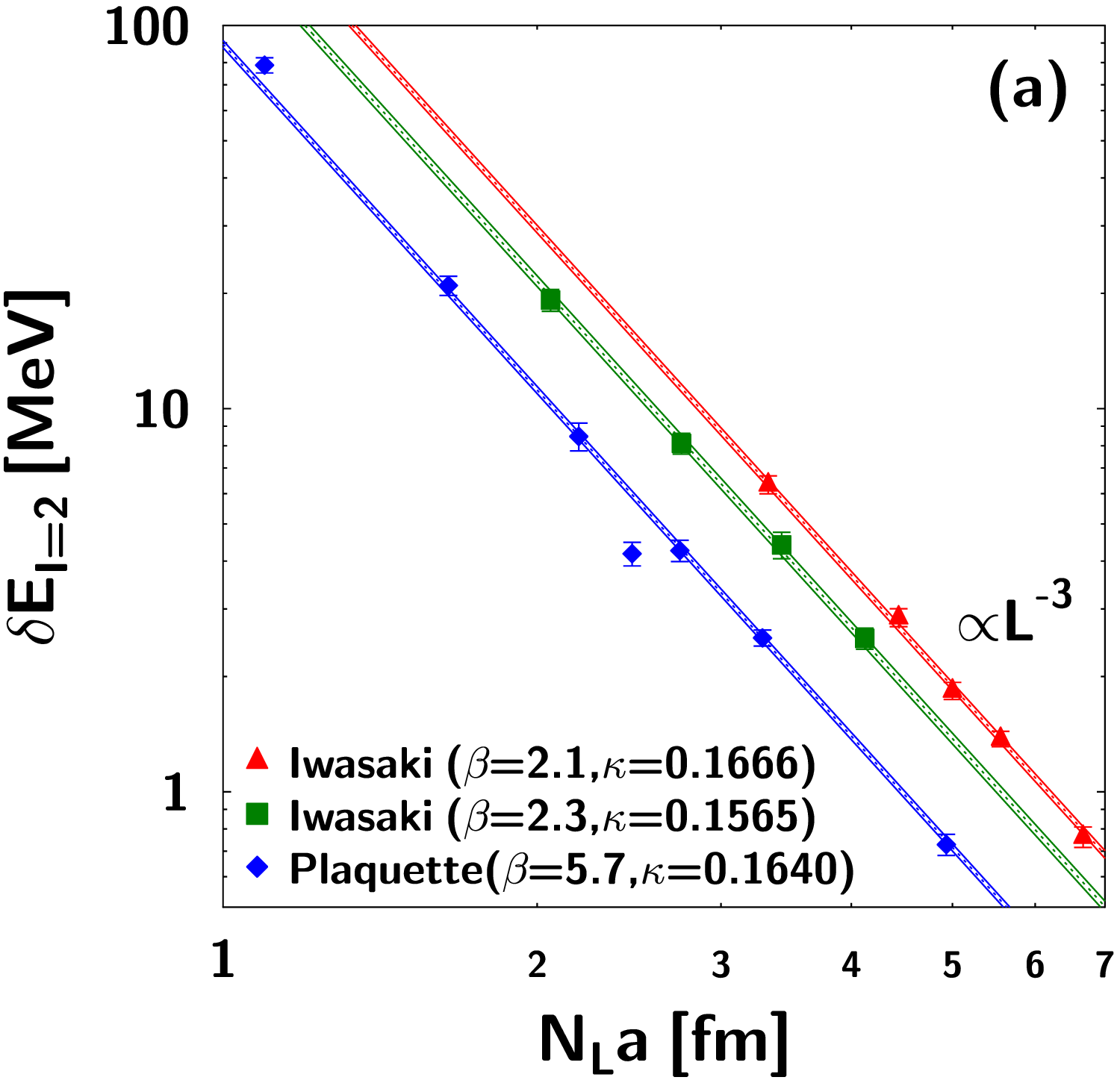} \ \ \ \ \ \ \ \ \ \ \ \
\includegraphics[scale=0.455]{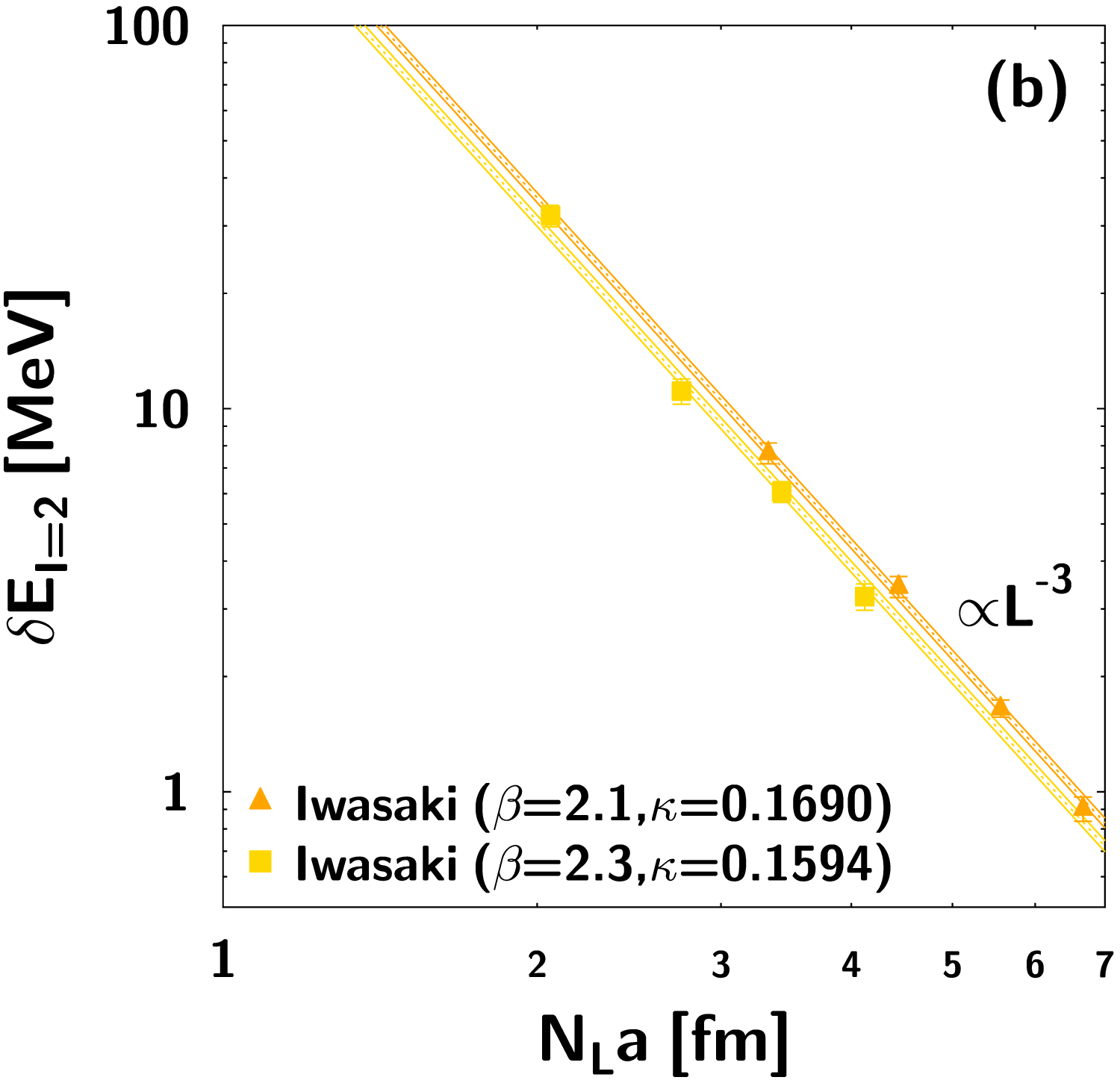}
\caption{The physical lattice size $L(=N_L a)$ dependence of the energy shifts $\delta E_{I=2}$. 
In left figure (a) (right figure (b)), the values of $m_{\pi}/m_{\rho}$ ($m_{\pi}$) are fixed. 
In both figures the lines are proportional to $L^{-3}$. 
In left figure (a), the three symbols are on the different $L^{-3}$ lines, 
though they have the same values of $m_{\pi}/m_{\rho}$. 
In right figure (b), the two symbols which have the same values of $m_{\pi}$ are almost on the same $L^{-3}$ line, 
though they have the different values of $m_{\pi}/m_{\rho}$. }
\label{EE_mm_m}
\end{center}
\end{figure}

\begin{figure}
\begin{center}
\includegraphics[scale=0.455]{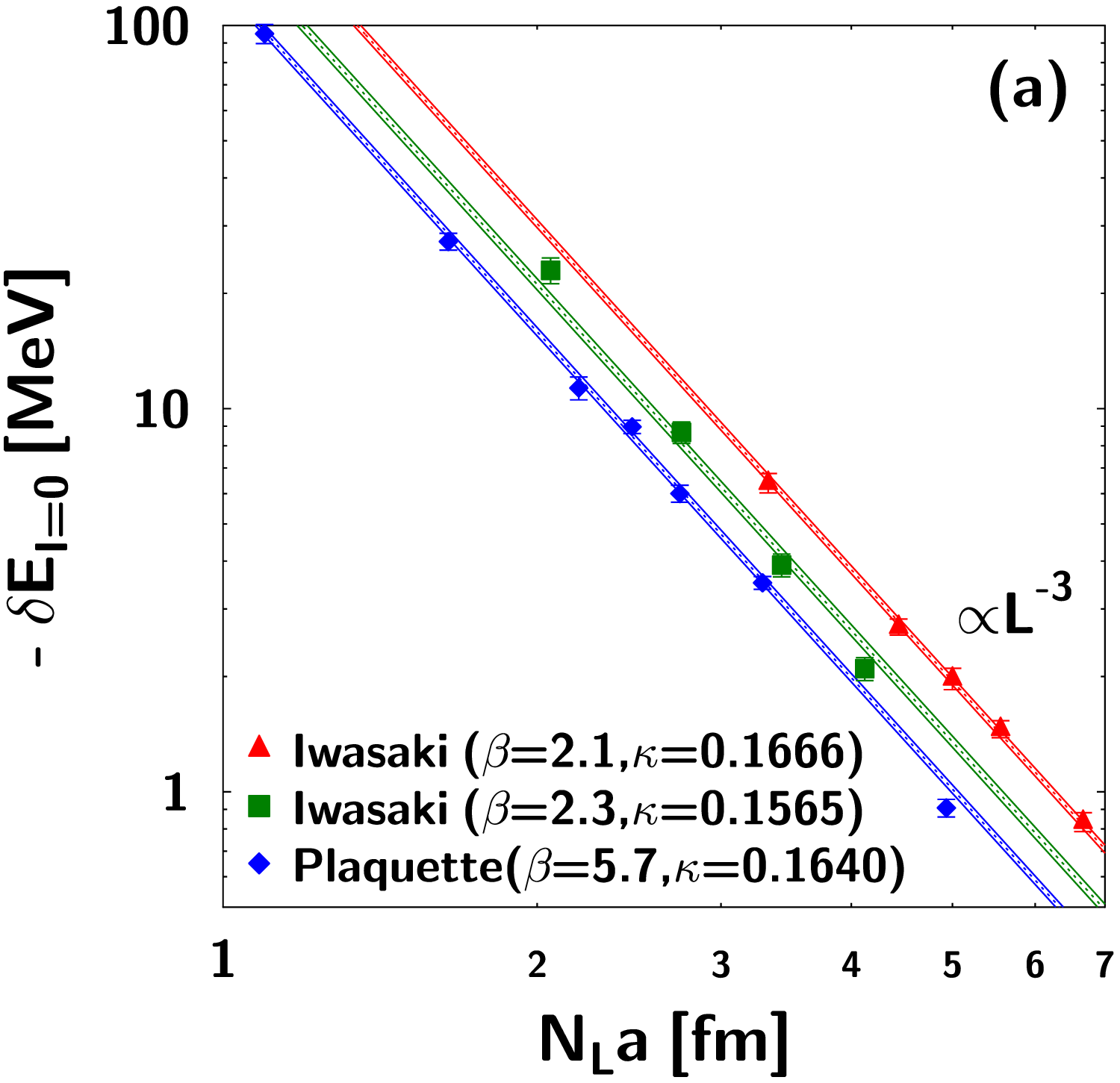} \ \ \ \ \ \ \ \ \ \ \ \
\includegraphics[scale=0.455]{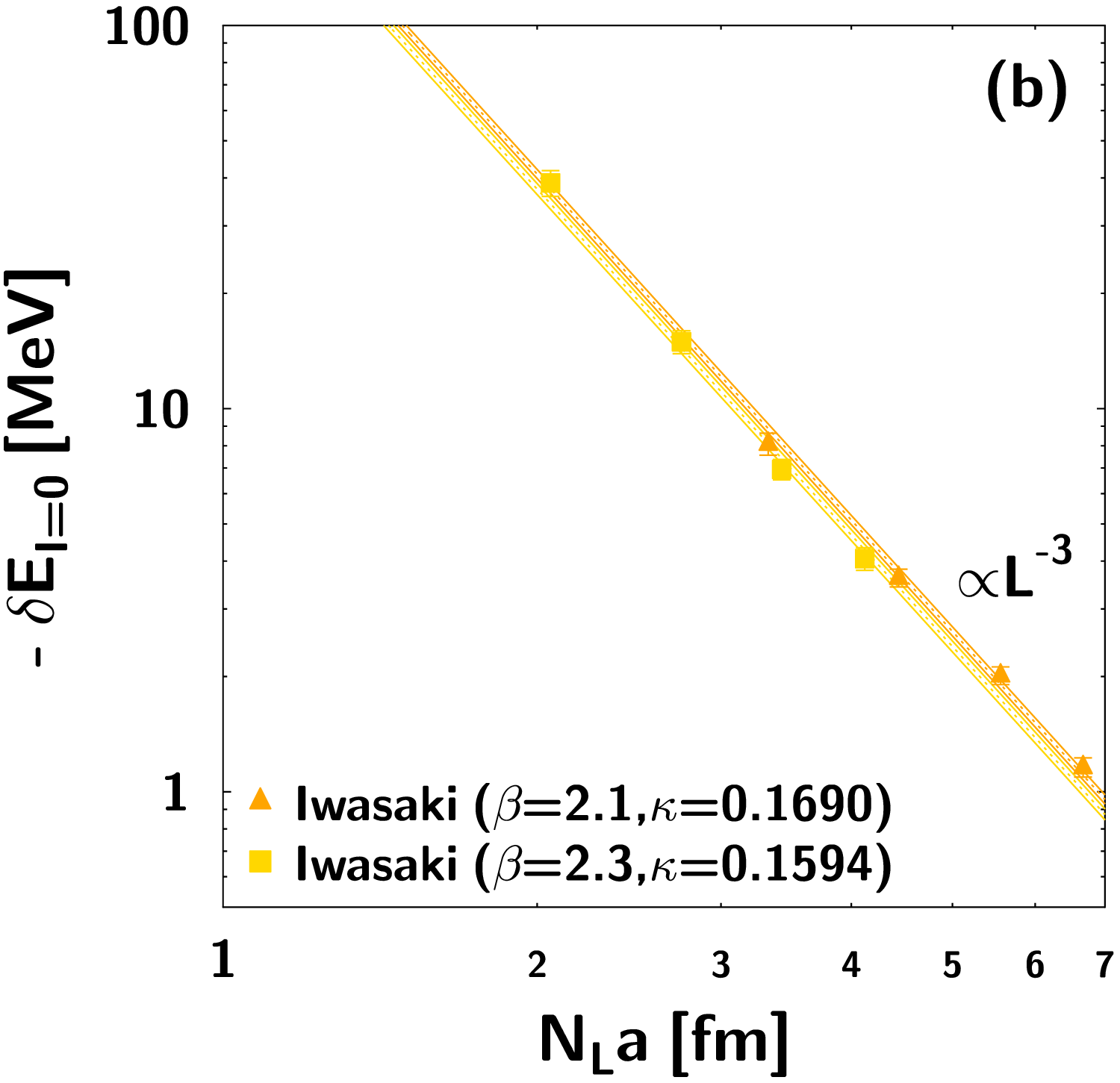}
\caption{The physical lattice size $L(=N_L a)$ dependence of the energy shifts $\delta E_{I=0}$. 
In left figure (a) (right figure (b)), the values of $m_{\pi}/m_{\rho}$ ($m_{\pi}$) are fixed. 
In both figures the lines are proportional to $L^{-3}$. 
In left figure (a), the three symbols are on the different $L^{-3}$ lines, 
though they have the same values of $m_{\pi}/m_{\rho}$. 
In right figure (b), the two symbols which have the same values of $m_{\pi}$ are almost on the same $L^{-3}$ line, 
though they have the different values of $m_{\pi}/m_{\rho}$. }
\label{mm_m}
\end{center}
\end{figure}

\section{Results}
In the work by Alford et al.\ \cite{Jaffe}, 
they observed that the $I=2$ channel is clearly the scattering state 
but found the possibility that the $I=0$ channel is the four-quark bound state.
However to get the conclusive result for the four-quark bound state in the $I=0$ channel, 
further calculations on a larger lattice with finer lattice spacing are required.

We calculate the four-quark correlators on a larger lattice with $N_f=2$ Wilson fermion 
using configurations which are produced with both plaquette and Iwasaki gauge actions.
The lattice parameters are shown in Tables \ref{pa1} and \ref{pa2}.
We impose the Dirichlet boundary condition in the temporal direction on the quark fields.

The results for $I=2$ channel are shown in Fig.\ \ref{EE_mm_m}.
In Fig.\ \ref{EE_mm_m} (a), 
the quark masses are set as the values of the ratio of pion mass to rho meson mass, $m_{\pi}/m_{\rho} \sim 0.74$ 
become close to those in Ref.\ \cite{Jaffe} (See Table \ref{para}).
In spite of the same values of $m_{\pi}/m_{\rho}$, the values of $m_{\pi}$ are different among them, 
which comes from the lattice artifact due to the large lattice spacings. 
In Fig.\ \ref{EE_mm_m} (b), 
the bound energies are obtained under the same values of $m_{\pi} \sim 370$ MeV.
We can see that in all cases of Fig.\ref{EE_mm_m} (a) and (b) the symbols are on the $L^{-3}$ lines, 
which is the same as that in Ref.\ \cite{Jaffe}.
On the other hand the results for $I=0$ channel are shown in Fig.\ \ref{mm_m}.
In Fig.\ \ref{mm_m} (a), the quark masses are set as the values of the ratio of $m_{\pi}/m_{\rho}$ 
become close to those in Ref.\ \cite{Jaffe} (See Table \ref{para}).
In Fig.\ \ref{mm_m} (b), 
the bound energies are obtained under the same values of $m_{\pi} \sim 370$ MeV.
Again we find that in all cases of Figs.\ref{mm_m} (a) and (b) the symbols are on the $L^{-3}$ lines,
which is contradicted with the result of Ref.\ \cite{Jaffe}.
We can not observe the suggestion of the existence of the bound state in the $I=0$ channel.

In Fig.\ \ref{mm_m} (a) 
the three symbols which have the same values of $m_{\pi}/m_{\rho}$ are on the different $L^{-3}$ lines, 
but in Fig.\ \ref{mm_m} (b) 
the two symbols which have the same values of $m_{\pi}$ are almost on the same $L^{-3}$ line.
It implies that $\delta E_{I}$ depends strongly on $m_{\pi}$ rather than $m_{\pi}/m_{\rho}$, 
which is deduced from Eq.(\ref{amp}).
Since the scattering amplitude is proportional to the inversion of $m_{\pi}$,
the absolute value of $\delta E_{I}$ decreases with the increasing amount of $m_{\pi}$.

\begin{figure}
\begin{center}
\includegraphics[scale=0.455]{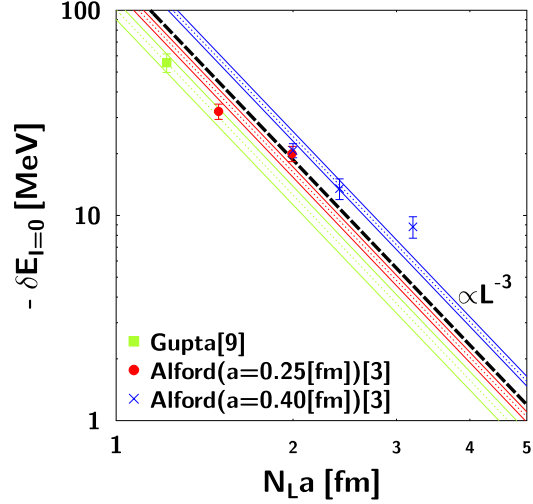}
\caption{The physical lattice size $L(=N_L a)$ dependence of the energy shifts $\delta E_{I=0}$. 
The lines are proportional to $L^{-3}$. 
The color difference corresponds to $m_{\pi}$ difference.
The green, red and blue solid lines stand for $m_{\pi} = 940, 840$ and $790$ MeV, respectively.
The black dashed line was drawn in Fig.5 of Ref.\cite{Jaffe}.}
\label{pri}
\end{center}
\end{figure}

In Ref.\ \cite{Jaffe} they insisted that 
they found the deviation from the $L^{-3}$ line (the black dashed line in Fig.\ \ref{pri}) 
in the physical lattice size dependence of energy shifts in the $I=0$ channel.
However we can give another explanation to it.
Because of the $m_{\pi}$ differences among data plots, 
we can draw the three different $L^{-3}$ lines (the green, red and blue solid lines) 
instead of one $L^{-3}$ line (the black dashed line) in Fig.\ \ref{pri}.
It does not suggest that the symbols deviate from the $L^{-3}$ line (the black dashed line), 
but suggests that the symbols are on the three different $L^{-3}$ lines, respectively.
It means that there are no bound states in the $I=0$ channel.

\section{Summary}
We have investigated a four-quark state as a bound state 
from the spatial lattice size dependence of the bound energies for a four-quark state. 
We found that 
the four-quark states in the cases of both $I=0$ and $I=2$ channels exist as no bound states, i.e., 
all bound energies for the four-quark states as a function of spatial lattice size $L$ are on the $L^{-3}$ lines, 
which is contradicted with the previous work \cite{Jaffe}.
The symbols which have the same values of $m_{\pi}/m_{\rho}$ are on the different $L^{-3}$ lines, 
on the other hand the symbols which have the same values of $m_{\pi}$ are almost on the same $L^{-3}$ line.
This results suggest that 
the absolute value of bound energy depends directly on $m_{\pi}$ rather than $m_{\pi}/m_{\rho}$.

This work is a staring point of study of four-quark states in the light scalar mesons.
To reach the conclusive results for the light scalar meson states, 
further investigation is needed; 
full QCD calculation with light quark masses on larger lattice size, 
application of four-quark interpolators, state-of-the-art action and so on.

\begin{table}[b]
\caption{Lattice parameters. 
$\kappa$ is the hopping parameter. 
We adjust parameters, $\beta$ and $\kappa$ to fix the values of $m_{\pi}/m_{\rho} \sim 0.74$.  
See Table \protect \ref{para}. }
\begin{center}
\begin{tabular}{c|c|c|c|c|cc|c}\hline
 \ \ \ Gauge action \ \ \ & $\beta$ & $a \ \scriptstyle{\mathrm{[fm]}}$ & $\kappa$ &$N_{L}^3\times N_{T}$ & \# Conf. \\\hline
Iwasaki    & 2.100 & 0.278(3) \cite{Iwa}    & 0.1666 & $12^3 \times  20\ $ & 150 \\
                &          &                                   &             & $16^3 \times  20\ $ & 120 \\
                &          &                                   &             & $18^3 \times  20\ $ &  78 \\
                &          &                                   &             & $20^3 \times  20\ $ & 120 \\
                &          &                                   &             & $24^3 \times  20\ $ &  78 \\\hline
Iwasaki    & 2.300 &  0.172(4) \cite{Iwa}   & 0.1565 & $12^3 \times  20\ $ & 120 \\ 
                &          &                                   &             & $16^3 \times  20\ $ &  60 \\ 
                &          &                                   &             & $20^3 \times  20\ $ &  60 \\
                &          &                                   &             & $24^3 \times  20\ $ &  54 \\\hline
Plaquette & 5.700 & 0.140(4) \cite{Fuku} & 0.1640 & $ \ 8^3 \times  20$ & 1008 \\ 
                &          &                                   &             & $12^3 \times  20\ $ & 240  \\ 
                &          &                                   &             & $16^3 \times  20\ $ & 102  \\
                &          &                                   &             & $18^3 \times  20\ $ & 210  \\
                &          &                                   &             & $20^3 \times  20\ $ & 192 \\
                &          &                                   &             & $24^3 \times  20\ $ & 300 \\
                &          &                                   &             & $36^3 \times  20\ $ & 66 \\\hline
\end{tabular}
\end{center}
\label{pa1}
\end{table}

\begin{table}[b]
\caption{Lattice parameters.  
We adjust parameters, $\beta$ and $\kappa$ to fix the values of $m_{\pi} \sim 370$ MeV.  
See   Table \protect \ref{para}.}
\begin{center}
\begin{tabular}{c|c|c|c|c|cc|c}\hline
 \ \ \ Gauge action \ \ \ & $\beta$ & $a \ \scriptstyle{\mathrm{[fm]}}$ & $\kappa$ &$N_{L}^3\times N_{T}$ & \# Conf. \\\hline
Iwasaki & 2.100 & 0.278(3) \cite{Iwa} & 0.1690  & $12^3 \times  20\ $ & 150   \\
             &           &               &               & $16^3 \times  20\ $ & 120  \\
             &           &               &               & $20^3 \times  20\ $ & 120 \\
             &           &               &               & $24^3 \times  20\ $ &  60 \\\hline
Iwasaki & 2.300 & 0.172(4) \cite{Iwa} & 0.1594  & $12^3 \times  20\ $ &  300 \\ 
             &           &               &               & $16^3 \times  20\ $ &  180 \\ 
             &           &               &               & $20^3 \times  20\ $ &  180 \\
             &           &               &               & $24^3 \times  20\ $ &   60 \\\hline
\end{tabular}
\end{center}
\label{pa2}
\end{table}

\begin{table}
\caption{Lattice parameters in Ref.\ \cite{Jaffe}. 
They adjust parameters, $\beta$ and $\kappa$ to fix the values of $m_{\pi}/m_{\rho} \sim 0.74$.  
See Table \protect \ref{para}. }
\begin{center}
\begin{tabular}{c|c|c|c|c|c|cc|c}\hline
Group & Gauge action &$\beta$ & $a \ \scriptstyle{\mathrm{[fm]}}$ & $\kappa$ &$N_{L}^3\times N_{T}$ & \# Conf. \\\hline
 Gupta et al. \cite{Gupta} &Plaquette               & 6.000 & 0.0762(8)$\!\!$ \cite{Davies} & 0.1540 & $16^3 \times  80\ $ & 35 \\ \hline
 Fukugita et al. \cite{Fuku} &Plaquette             & 5.700 & 0.162(6) \ \cite{Davies} & 0.1640 & $12^3 \times  20\ $ & 70 \\ \hline
 Alford et al. &L$\ddot{\rm{u}}$scher \& Weisz & 1.719 &0.249(5)    & -----      & $\ 6^3 \times  40$   & ----- \\ 
                    &                                                   &           &                 &             & $\ 8^3 \times  40$   & ----- \\\hline
 Alford et al. &L$\ddot{\rm{u}}$scher \& Weisz & 1.157 & 0.400(4)   &   -----    & $\ 5^3 \times  40$   & ----- \\ 
                    &                                                   &           &                 &             & $\ 6^3 \times  40$   & ----- \\ 
                    &                                                   &           &                 &             & $\ 8^3 \times  40$   & ----- \\\hline
\end{tabular}
\end{center}
\label{pa0}
\end{table}

\begin{table}
\caption{The values of $m_{\pi}/m_{\rho}$ and $m_{\pi}$. 
In the upper table results of Ref.\ \cite{Jaffe} are shown. 
In the lower table results of our calculation are shown.}
\begin{center}
\begin{tabular}{c|c|c}\hline
 data  & $m_{\pi}/m_{\rho}$ &$m_{\pi} \ \scriptstyle{\mathrm{[MeV]}}$ \\\hline
 Gupta                                                                & -----        & 940(30)    \\ 
 Fukugita                                                            & 0.740(8) & 620(90)    \\ 
 Alford ($a=0.25$ [fm])                                       & 0.756(5) & 840(11)   \\ 
 Alford ($a=0.40$ [fm])                                       & 0.756(4) & 790(6) \ \       \\\hline\hline
 Iwasaki ($\beta=2.1, \kappa=0.1666$)             & 0.736(3) & 451(6) \ \    \\ 
 Iwasaki ($\beta=2.3, \kappa=0.1565$)             & 0.746(6) & 593(16)    \\ 
 Plaquette ($\beta=5.7, \kappa=0.1640$) \ \ \   & 0.748(4) & 754(12)      \\\hline
 Iwasaki ($\beta=2.1, \kappa=0.1690$)             & 0.646(4) & 368(5) \ \       \\
 Iwasaki ($\beta=2.3, \kappa=0.1594$)             & 0.545(7) & 374(11)    \\\hline
\end{tabular}
\end{center}
\label{para}
\end{table}

\section*{Acknowledgments}
This work is supported in part by 
Nagoya University Global COE Program (G07).
Numerical calculations were performed on 
the cluster system "$\varphi$" at KMI, Nagoya University.
Grant-in-Aid for Young Scientists (B) (22740156),
Grant-in-Aid for Scientific Research (S) (22224003) and the Kurata Memorial Hitachi
Science and Technology Foundation.

\end{document}